\documentclass[11pt]{article}

\pdfoutput=1

\usepackage[T1]{fontenc}
\usepackage[latin9]{inputenc}
\usepackage[a4paper]{geometry}
\usepackage[active]{srcltx}
\usepackage{amsmath}
\usepackage{amssymb}
\usepackage{esint}

\makeatletter

\pdfoutput=1 

\usepackage{jheppub}

\usepackage{etoolbox}
    
    \patchcmd{\maketitle}{\@fpheader}{}{}{}

\usepackage[T1]{fontenc}
\usepackage{ae,aecompl}

\usepackage{amsfonts}

\usepackage{bbm}

\title{\boldmath Electrically charged black hole on AdS$_3$: scale invariance and the Smarr formula}

\author[a,b]{Cristián Erices,}
\author[a]{Oscar Fuentealba,}
\author[a]{Miguel Riquelme}

\affiliation[a]{Centro de Estudios Cient\'{i}ficos (CECs), Av. Arturo Prat 514, Valdivia, Chile.}
\affiliation[b]{Department of Physics, National Technical University of Athens, Zografou Campus GR 157 73, Athens, Greece.}

\emailAdd{erices@cecs.cl}
\emailAdd{fuentealba@cecs.cl}
\emailAdd{riquelme@cecs.cl}

\preprint{}

\abstract{The Einstein-Maxwell theory with negative cosmological constant in three spacetime dimensions is considered. It is shown that the Smarr relation for the electrically charged BTZ black hole emerges from two different approaches based on the scaling symmetry of the asymptotic behaviour of the fields at infinity. In the first approach, we prove that the conservation law associated to the scale invariance of the action for a class of stationary and circularly symmetric configurations, allows to obtain the Smarr formula as long as a special set of holographic boundary conditions is satisfied. This particular set is singled out making the integrability conditions for the energy compatible with the scale invariance of the reduced action. In the second approach, it is explicitly shown that the Smarr formula is recovered through the Euler theorem for homogeneous functions, provided the same set of holographic boundary conditions is fulfilled.}

\makeatother

\begin{document}
\maketitle \flushbottom

\section{Introduction\label{Introduction}}

Since the early stage of the thermodynamical description of black
holes, the Smarr formula \cite{Smarr:1972kt} has been an intensive
subject of study as an analogous of the Euler equation for black hole
mechanics. This formula can be understood as an integrated form of
the first law under certain homogeneity assumptions for the extensive
variables. Specifically, this relation states the energy as a bilinear
form of the global charges of the black hole along with their corresponding
chemical potentials, as long as the entropy is a homogeneous function
of a definite degree in the conserved charges. In the case of three-dimensional
Einstein gravity on AdS$_{3}$, the BTZ black hole \cite{Banados:1992wn},
\cite{Banados:1992gq} naturally satisfies this requirement. Indeed,
the entropy of the BTZ black hole can be written in terms of the global
charges through the Cardy formula \cite{Strominger:1997eq};
\begin{equation}
S=2\pi\sqrt{\frac{c}{12}\left(Ml+J\right)}+2\pi\sqrt{\frac{c}{12}\left(Ml-J\right)}\,,\label{eq:CardyBTZ}
\end{equation}
where $c=\frac{3l}{2G}$ stands for the Brown-Henneaux central charge
\cite{Brown:1986nw}. From \eqref{eq:CardyBTZ} is evident that the
entropy is a homogeneous function of degree $\frac{1}{2}$ in $\left(M,J\right)$.
Hence, by direct application of the Euler theorem, it is possible
to write the energy as a Smarr relation \cite{Cai:1996df}, \cite{Clement:2003sr};
\begin{equation}
M=\frac{1}{2}TS+\Omega J\,.
\end{equation}

In the Einstein-Maxwell theory with negative cosmological constant
in three dimension the situation is rather different. As shown in
\cite{Perez:2015kea}, \cite{Perez:2015jxn}, the energy spectrum
of the electrically charged rotating BTZ black hole \cite{Clement:1993kc},
\cite{Martinez:1999qi} is highly sensitive to the choice of boundary
conditions,\footnote{It is worth to point out that we refer to boundary conditions to conditions
that are held fixed at the boundary, while we mean by asymptotic conditions
to the asymptotic behaviour of the fields at infinity.} what makes its thermodynamical description a very subtle problem.
For instance, for the simplest choice of boundary conditions the energy
spectrum turns out to be unbounded from below \cite{Martinez:1999qi}.
Indeed, by considering the same set of boundary conditions, the following
formula of the energy holds
\begin{equation}
M=\frac{1}{2}TS+\Omega J+\frac{1}{2}\Phi_{e}Q_{e}+\frac{1}{8\pi}\left(1-l^{2}\Omega^{2}\right)Q_{e}^{2}\,,\label{eq:SmarrSpoiled}
\end{equation}
which is certainly not a Smarr relation. As mentioned in \cite{Bravo-Gaete:2015iwa},
this is because of the logarithmic contributions of the gauge potential,
which spoils the homogeneity property of the extensive quantities,
in the sense that entropy is no longer a homogeneous function with
a definite degree in the conserved charges. In what follows, we show
that it possible to recover the aforementioned homogeneity property
by considering the asymptotic conditions of the Einstein-Maxwell theory
on AdS$_{3}$ introduced in \cite{Perez:2015kea}, \cite{Perez:2015jxn}
endowed with an appropriate set of boundary conditions.

The aim of this work is to show how the Smarr formula for the charged
BTZ black hole emerges through two different approaches. Both of them
are based on the preservation of the fall-off of the fields at infinity,
given in \cite{Perez:2015kea}, under a specific set of scale transformations
that leaves the reduced action principle invariant for a wide class
of configurations. In particular, we will use the method developed
in \cite{Banados:2005hm} that recovers the Smarr formula for three-dimensional
hairy black holes from a radial conservation law related to a scale
invariance of the reduced action. However, the assumptions considered
in this method are that the matter fields must be finite at the event
horizon and vanish at infinity, where the latter is clearly not satisfied
by the charged BTZ black hole because of the presence of the logarithmic
terms. In spite of that, we will show herein that this method can
still be applied in the case of the Einstein-Maxwell theory on AdS$_{3}$
by implementing the asymptotic conditions proposed in \cite{Perez:2015kea},
\cite{Perez:2015jxn}. As consequence, it can be proved that the Smarr
formula for the charged BTZ black hole holds as long as a special
set of holographic boundary conditions is satisfied. 

The next section is dedicated to a brief review of the main results
found in \cite{Perez:2015kea}, related to the global charges and
their integrability conditions for stationary and circularly symmetric
configurations in the Einstein-Maxwell theory on AdS$_{3}$. In section
\ref{ScaleInv}, we prove that the conservation law associated to
the scale invariance of the action for the aforementioned class of
configurations, allows to obtain the Smarr formula as long as a special
set of holographic boundary conditions is satisfied. This particular
set is singled out by requiring compatibility of the integrability
conditions for the energy with the scale invariance of the reduced
action principle. In section \ref{Euler}, it is shown that the same
set of holographic boundary conditions ensures the right homogeneous
transformation laws for the extensive quantities of the black hole
under scaling transformations, allowing to recover the Smarr formula
through the Euler theorem along the lines of its original derivation
for the Kerr-Newman black hole in \cite{Smarr:1972kt}. We conclude
with some ending remarks in section \ref{Concludings}.

\section{A review on the Einstein-Maxwell theory on AdS$_{3}$ and global
charges}

This section is devoted to a brief review of the results found in
\cite{Perez:2015kea}. It is shown the reduced action principle of
the Einstein-Maxwell theory on AdS$_{3}$ in a canonical form for
stationary and circularly symmetric configurations, the variation
of the global charges and their appropriate integrability conditions.

\subsection{Action principle for stationary and circularly symmetric configurations}

The action of the Einstein-Maxwell theory with negative cosmological
constant in three spacetime dimensions is given by
\begin{equation}
I_{EM}=\int d^{3}x\sqrt{-g}\left[\frac{1}{2\kappa}\left(R-2\Lambda\right)-\frac{1}{4}F_{\alpha\beta}F^{\alpha\beta}\right]\ .\label{eq:Einstein-Maxwell action}
\end{equation}
Here the Newton constant $G$ and the AdS radius $l$ are defined
through $\kappa=8\pi G$ and $\Lambda=-l^{-2}$, respectively.

We consider stationary and circularly symmetric spacetimes, which
describe a wide class of configurations already reported in the literature
\cite{Banados:1992wn}, \cite{Clement:1993kc}, \cite{Martinez:1999qi},
\cite{Deser:1985pk}, \cite{Gott:1986bp}, \cite{Peldan:1992mp},
\cite{Kamata:1995zu}, \cite{Chan:1995uh}, \cite{Clement:1995zt},
\cite{Hirschmann:1995he}, \cite{Cataldo:1996ue}, \cite{Kamata:1996vg},
\cite{Cataldo:1999fh}, \cite{Cataldo:2002fh}, \cite{Dias:2002ps},
\cite{Cataldo:2004uw}, \cite{Matyjasek:2004pg}, \cite{Garcia-Diaz:2013baa},
\cite{AyonBeato:2004qr}. The line element for this family of solutions
is given by
\begin{equation}
ds^{2}=-\mathcal{N}\left(r\right)^{2}\mathcal{F}\left(r\right)^{2}dt^{2}+\frac{dr^{2}}{\mathcal{F}\left(r\right)^{2}}+\mathcal{R}\left(r\right)^{2}\left(\mathcal{N}^{\phi}\left(r\right)dt+d\phi\right)^{2}\ ,\label{Stationary metric}
\end{equation}
where the gauge field is chosen as
\begin{equation}
A=\mathcal{A}_{t}\left(r\right)dt+\mathcal{A}_{\phi}\left(r\right)d\phi\ .\label{Gauge field}
\end{equation}
The reduced action principle in the canonical form can be obtained
by replacing the class of configurations described by \eqref{Stationary metric}
and \eqref{Gauge field} in \eqref{eq:Einstein-Maxwell action}, which
reads
\begin{equation}
I=-2\pi\left(t_{2}-t_{1}\right)\int dr\left(\mathcal{NH}+\mathcal{N^{\phi}}\mathcal{H}_{\phi}+\mathcal{A}_{t}\mathcal{G}\right)+B\ ,\label{reduced action}
\end{equation}
where the boundary term $B$ must be added in order to have a well-defined
variational principle. The surface deformation generators $\mathcal{H}$,
$\mathcal{H_{\phi}}$ and the generator of gauge transformations $\mathcal{G}$,
acquire the following form
\begin{align}
\mathcal{H} & =-\frac{\mathcal{R}}{\kappa l^{2}}+4\kappa\mathcal{R}\left(\pi^{r\phi}\right)^{2}+\frac{\left(p^{r}\right)^{2}}{2\mathcal{R}}+\frac{\mathcal{F}^{2}\left(\mathcal{A}_{\phi}^{\prime}\right)^{2}}{2\mathcal{R}}+\frac{\left(\mathcal{F}^{2}\right)^{\prime}\mathcal{R}^{\prime}}{2\kappa}+\frac{\mathcal{F}^{2}\mathcal{R}^{\prime\prime}}{\kappa}\ ,\label{H}\\
\mathcal{H}_{\phi} & =-p^{r}\mathcal{A}_{\phi}^{\prime}-2\left(\mathcal{R}^{2}\pi^{r\phi}\right)^{\prime}\ ,\label{Hphi}\\
\mathcal{G} & \mathcal{=}-\partial_{r}p^{r}\ ,\label{G}
\end{align}
where $\mathcal{N}$, $\mathcal{N}^{\phi}$ and $\mathcal{A}_{t}$
stand for their corresponding Lagrange multipliers. The only nonvanishing
components of the momenta $\pi^{ij}$ and $p^{i}$ are explicitly
given by
\begin{equation}
\pi^{r\phi}=-\frac{\mathcal{N}^{\phi\prime}\mathcal{R}}{4\kappa\mathcal{N}}\ \ ;\ \ p^{r}=\frac{\mathcal{R}}{\mathcal{N}}\left(\mathcal{A}_{\phi}^{\prime}\mathcal{N}^{\phi}-\mathcal{A}_{t}^{\prime}\right)\ .\label{pr}
\end{equation}
Note that prime denotes derivative with respect to $r$. 

\subsection{Global charges and integrability conditions\label{GlobalCharges}}

Hereafter, it is considered that the class of configurations that
we are dealing with are given by asymptotically AdS$_{3}$ spacetimes
with the following behaviour at infinity\footnote{The dots ``$\cdots$'' stand for subleading terms that can be consistently
gauged away because they do not appear neither in the global charges
nor in the gauge transformations of the dynamical fields \cite{Benguria:1976in}. }
\begin{align}
\mathcal{R}^{2} & =r^{2}-\frac{\kappa l^{2}}{\pi}\left[h_{\mathcal{R}}\log\left(\frac{r}{l}\right)-\frac{f_{\mathcal{R}}}{2}\right]+\cdots\nonumber \\
\mathcal{F}^{2} & =\frac{r^{2}}{l^{2}}-\frac{\kappa}{\pi}\left[\left(2h_{\mathcal{R}}+\frac{1}{4\pi}\left(q_{t}^{2}+q_{\phi}^{2}\right)\right)\log\left(\frac{r}{l}\right)+f_{\mathcal{F}}\right]+\cdots\nonumber \\
\mathcal{N}^{\phi} & =N_{\infty}^{\phi}+\frac{\kappa}{2\pi}N_{\infty}\left[\frac{l}{2\pi}q_{t}q_{\phi}\log\left(\frac{r}{l}\right)-j\right]\frac{1}{r^{2}}+\cdots\label{fall-off}\\
\mathcal{N} & =N_{\infty}+\cdots\nonumber \\
\mathcal{A}_{t} & =-\frac{1}{2\pi}\left(q_{t}N_{\infty}+q_{\phi}lN_{\infty}^{\phi}\right)\log\left(\frac{r}{l}\right)+N_{\infty}^{\phi}\varphi_{\phi}+N_{\infty}\frac{\varphi_{t}}{l}-\Phi+\cdots\nonumber \\
\mathcal{A}_{\phi} & =-\frac{q_{\phi}l}{2\pi}\log\left(\frac{r}{l}\right)+\varphi_{\phi}+\cdots\nonumber 
\end{align}
which was proposed in \cite{Perez:2015kea}. Here the coefficients
$h_{\mathcal{R}}$, $f_{\mathcal{R}}$, $f_{\mathcal{F}}$, $j$,
$\varphi_{t}$, $\varphi_{\phi}$, $q_{t}$, $q_{\phi}$ are constants,
which are assumed to vary in the phase space, while $N_{\infty}$,
$N_{\infty}^{\phi}$ and $\Phi$ correspond to arbitrary constants
without variation, which are kept fixed at the boundary. 

Following the canonical approach given in \cite{Regge:1974zd}, as
shown in \cite{Perez:2015kea}, the variation of the energy for the
class of configurations \eqref{Stationary metric}, \eqref{Gauge field}
endowed with the fall-off for the fields \eqref{fall-off}, is given
by
\begin{equation}
\delta M=\delta\left[f_{\mathcal{R}}+f_{\mathcal{F}}+h_{\mathcal{R}}+\frac{1}{l}q_{\phi}\varphi_{\phi}\right]-\frac{1}{l}\varphi_{\mu}\delta q^{\mu}\ ,\label{generic dM}
\end{equation}
where $\varphi_{\mu}=\left(l^{-1}\varphi_{t},\varphi_{\phi}\right)$
and $q_{\mu}=\left(l^{-1}q_{t},q_{\phi}\right)$ are assumed to be
Lorentz covariant vectors, whose indices are raised and lowered by
the flat (conformal) boundary metric $\eta_{\mu\nu}=diag(-l^{-2},1)$.
The rest of the global charges; angular momentum $J$ and electric
charge $Q_{e}$, can be directly integrated, and they read
\begin{align}
J & =j+\frac{l}{4\pi}q_{t}q_{\phi}-q_{t}\varphi_{\phi}\ ,\label{eq:}\\
Q_{e} & =q_{t}\;.\label{Electric charge}
\end{align}

As explained in \cite{Perez:2015kea}, \eqref{generic dM} yields
a nontrivial integrability condition for $\varphi_{\mu}$ and $q_{\mu}$.
The integrability for the energy is ensured by the condition $\delta^{2}M=0$,
which is satisfied provided
\begin{equation}
\varphi_{\mu}=-\frac{\delta\mathcal{V}}{\delta q^{\mu}}\ .\label{definition NU}
\end{equation}
where $\mathcal{V}=\mathcal{V}\left(q^{\mu}\right)$ is an arbitrary
function of $q_{t}$ and $q_{\phi}$. In consequence, the energy and
the angular momentum are then given by

\begin{eqnarray}
M & = & f_{\mathcal{R}}+f_{\mathcal{F}}+h_{\mathcal{R}}+\frac{1}{l}\left(\mathcal{V}-q_{\phi}\frac{\delta\mathcal{V}}{\delta q_{\phi}}\right)\ ,\label{generic M}\\
J & = & j+\frac{l}{4\pi}q_{t}q_{\phi}+q_{t}\frac{\delta\mathcal{V}}{\delta q_{\phi}}\,.\label{generic J}
\end{eqnarray}

In sum, the global charges are determined by the function $\mathcal{V}$
that describes the set of boundary conditions compatible with the
integrability of the energy.

\section{Scale invariance and radial conservation law \label{ScaleInv}}

In this section, we will make use of the approach given in \cite{Banados:2005hm},
where the Smarr formula for three-dimensional hairy black holes is
recovered from a radial conservation law associated to a scale invariance
of the reduced action. The assumptions considered in this approach
are that the matter fields must be finite at the event horizon and
vanish at infinity, which due to the presence of the logarithmic terms
is clearly not satisfied by the charged BTZ black hole. Nonetheless,
we will show that this method can still be applied in the case of
the Einstein-Maxwell theory on AdS$_{3}$ by implementing the asymptotic
conditions proposed in \cite{Perez:2015kea}, \cite{Perez:2015jxn}.

In this case, it is possible to prove that the reduced action principle
given in \eqref{reduced action} is invariant under the following
set of transformations
\begin{equation}
\begin{array}{cccccc}
\bar{\mathcal{R}}\left(\bar{r}\right)=\lambda\mathcal{R}\left(r\right)\,, &  &  &  &  & \bar{\mathcal{N}}\left(\bar{r}\right)=\lambda^{-2}\mathcal{N}\left(r\right)\,,\\
\\
\bar{\mathcal{F}}\left(\bar{r}\right)^{2}=\lambda^{2}\mathcal{F}\left(r\right)^{2}\,, &  &  &  &  & \bar{\mathcal{N}}^{\phi}\left(\bar{r}\right)=\lambda^{-2}\mathcal{N}^{\phi}\left(r\right)\,,\\
\\
\bar{\mathcal{A}}_{\phi}\left(\bar{r}\right)=\lambda\mathcal{A}_{\phi}\left(r\right)\,, &  &  &  &  & \bar{\mathcal{A}_{t}}\left(\bar{r}\right)=\lambda^{-1}\mathcal{A}_{t}\left(r\right)\,,\\
\\
\bar{p}^{r}\left(\bar{r}\right)=\lambda p^{r}\left(r\right)\,, &  &  &  &  & \bar{\pi}^{r\phi}\left(\bar{r}\right)=\pi^{r\phi}\left(r\right)\,,
\end{array}\label{eq:Scaling-transf}
\end{equation}
spanned by the scalings $\bar{r}=\lambda r$, $\bar{t}=t$ and $\bar{\phi}=\phi$,
where $\lambda$ is a positive constant. Note that similar scaling
symmetries were firstly observed in the matter-free case \cite{Horowitz:1999jd}
and in the context of three-dimensional hairy black holes in \cite{Banados:2005hm}.

A direct application of the Noether theorem, by considering the infinitesimal
transformation laws derived from \eqref{eq:Scaling-transf} on the
reduced action principle \eqref{reduced action}, yields the following
conserved quantity
\begin{eqnarray}
C\left(r\right) & = & 2\pi p^{r}(\mathcal{A}_{t}+\mathcal{N}^{\phi}\mathcal{A}_{\phi})+8\pi\mathcal{N}^{\phi}\mathcal{R}{}^{2}\pi^{r\phi}\nonumber \\
 &  & -\frac{2\pi\mathcal{F}{}^{2}\mathcal{N}\mathcal{A}_{\phi}\mathcal{A}'_{\phi}}{\mathcal{R}}+\frac{\pi\mathcal{N}\mathcal{R}(\mathcal{F}^{2})'}{\kappa}+\frac{2\pi\mathcal{F}{}^{2}\mathcal{R}\mathcal{N}'}{\kappa}-\frac{2\pi\mathcal{F}{}^{2}\mathcal{N}\mathcal{R}'}{\kappa}\label{eq:Noether-charge}\\
 &  & +\mathcal{N}\mathcal{R}\left(\frac{2\pi}{\kappa l^{2}}-\frac{\pi(p^{r})^{2}}{\mathcal{R}^{2}}-8\pi\kappa(\pi^{r\phi})^{2}+\frac{\pi\mathcal{F}{}^{2}(\mathcal{A}{}_{\phi}')^{2}}{\mathcal{R}^{2}}-\frac{\pi(\mathcal{F}^{2})'\mathcal{R}'}{\kappa\mathcal{R}}-\frac{2\pi\mathcal{F}{}^{2}\mathcal{N}'\mathcal{R}'}{\kappa\mathcal{N}\mathcal{R}}\right)r\,,\nonumber 
\end{eqnarray}
along the radial direction, i.e. $C'=0$, by virtue of the field equations.
We will explore whether it is possible to find a Smarr formula for
the charged rotating black hole \cite{Clement:1993kc}, \cite{Martinez:1999qi}
from the conserved quantity \eqref{eq:Noether-charge}. Thus, in the
particular case of the black hole solution with event horizon located
at $r_{+}$, $C(\infty)=C(r_{+})$. 

In what follows, we proceed to compute $C(r)$ at infinity by considering
the asymptotic behaviour of the fields given in section \ref{GlobalCharges},
and then at the event horizon by imposing appropriate regularity conditions.

\subsection{Conserved charge at infinity: holographic boundary conditions}

By considering the fall-off of the fields \eqref{fall-off}, the radial
conserved charge \eqref{eq:Noether-charge} at the asymptotic region
becomes
\begin{eqnarray}
C\left(\infty\right) & = & 2N_{\infty}\left[f_{\mathcal{F}}+f_{\mathcal{R}}+h_{\mathcal{R}}+\frac{l}{8\pi}\left(q_{\phi}^{2}-q_{t}^{2}\right)+\frac{1}{l}\left(q_{t}\varphi_{t}+q_{\phi}\varphi_{\phi}\right)\right]\nonumber \\
 &  & -2N_{\infty}^{\phi}\left(j+\frac{l}{4\pi}q_{t}q_{\phi}-q_{t}\varphi_{\phi}\right)-\Phi q_{t}\,,
\end{eqnarray}
recalling that $\varphi_{\mu}=-\frac{\delta\mathcal{V}}{\delta q^{\mu}}$. 

In order to determine the functional form of $\mathcal{V}$ some physically
reasonable criteria must be used. In this case, we will require compatibility
of the asymptotic conditions given in \eqref{fall-off} and the scale
invariance of the reduced action under transformations \eqref{eq:Scaling-transf},
which allows to find the explicit form of this function. In particular,
considering the scale transformations $\bar{A_{t}}\left(\bar{r}\right)=\lambda^{-1}A_{t}\left(r\right)$
and $\bar{A_{\phi}}\left(\bar{r}\right)=\lambda A_{\phi}\left(r\right)$
implies the following transformation rules for $\varphi_{\mu}$ and
$q_{\mu}$
\begin{equation}
\bar{\varphi}_{\mu}=\lambda\left(\varphi_{\mu}+\frac{lq_{\mu}}{2\pi}\log\left(\lambda\right)\right)\quad,\quad\bar{q}_{\mu}=\lambda q_{\mu}\;.\label{eq:Transphi}
\end{equation}
It must be highlighted that the transformations rules \eqref{eq:Transphi}
precisely coincide with the ones found in \cite{Perez:2015kea}, where
it was made use of a scaling symmetry that leaves the configuration
invariant and rescales the reduced action as $\bar{I}=\lambda^{2}I$. 

Compatibility of the transformation rules \eqref{eq:Transphi} under
scalings with the integrability condition for the energy \eqref{definition NU}
requires, up to an arbitrary integration constant without variation,
that the function $\mathcal{V}$ must obey the following differential
equation
\begin{equation}
q_{t}\frac{\partial\mathcal{V}}{\partial q_{t}}+q_{\phi}\frac{\partial\mathcal{V}}{\partial q_{\phi}}=2\mathcal{V}+\frac{l}{4\pi}\left(q_{t}^{2}-q_{\phi}^{2}\right)\,.\label{eq:Euler theorem2}
\end{equation}
The general solution of equation \eqref{eq:Euler theorem2} is given
by
\begin{equation}
\mathcal{V}=q_{t}^{2}F\left(\frac{q_{\phi}}{q_{t}}\right)+\frac{l}{8\pi}\left[q_{t}^{2}(\log[q_{t}^{2}]-1)-q_{\phi}^{2}(\log[q_{\phi}^{2}]-1)\right]\,,\label{eq:Holographic}
\end{equation}
where $F$ is an arbitrary function that describes a special set of
boundary conditions compatible with the scale invariance. Hereafter,
following \cite{Perez:2015kea}, we will refer them as ``holographic
boundary conditions''.\footnote{It is worth to note that requiring invariance of the holographic boundary
conditions under Lorentz symmetry singles out a very special set of
boundary conditions, which remarkably, has shown to be compatible
with the full conformal symmetry at infinity \cite{Perez:2015jxn}.} 

By using the integrability condition \eqref{definition NU} and holographic
boundary condition determined by \eqref{eq:Euler theorem2}, it is
found that $C(\infty)$ in terms of the global charges \eqref{generic dM},
\eqref{eq:} and \eqref{Electric charge} is given by
\begin{equation}
C\left(\infty\right)=2N_{\infty}M-2N_{\infty}^{\phi}J-\Phi Q_{e}\,.\label{eq:Cinf}
\end{equation}
In the following subsection we focus in the value of the Noether quantity
\eqref{eq:Noether-charge} in the case of the configurations that
possess an event horizon, that is the case of the charged BTZ black
hole, where it is mandatory to impose some regularity conditions in
order to ensure a well-defined Euclidean action principle (see e.g.
\cite{Bunster:2014mua}).

\subsection{Conserved charge at the event horizon: regularity conditions}

In order to evaluate the conserved charge \eqref{eq:Noether-charge}
at the event horizon, that for the class of configurations \eqref{Stationary metric}
is determined by $\mathcal{F}^{2}(r_{+})=0$, we have to consider
smooth configurations which must satisfy regularity of the Euclidean
geometry around the event horizon. These regularity conditions are
generically given by\textbf{
\begin{equation}
\mathcal{A}_{\tau}(r_{+})=0\quad,\quad\mathcal{N}^{\phi}(r_{+})=0\quad,\quad\mathcal{N}(r_{+})(\mathcal{F}^{2})'(r_{+})=4\pi\,.
\end{equation}
}Hence, making use of the constraint $\mathcal{H}=0$, \eqref{eq:Noether-charge}
at the event horizon becomes
\begin{equation}
C(r_{+})=\frac{4\pi^{2}}{\kappa}\mathcal{R}(r_{+})=S\,,\label{eq:Crp}
\end{equation}
which turns out to be the well-known Bekenstein-Hawking entropy of
the charged BTZ black hole, i.e., $S=\frac{A}{4G}$ recalling that
$\kappa=8\pi G$. 

Finally, the Smarr formula naturally emerges as consequence of the
equality $C\left(r_{+}\right)=C\left(\infty\right)$, and it is given
by\footnote{Similar relations for the entropy of three-dimensional black holes
and cosmological configurations as a bilinear combination of the global
charges along with their corresponding chemical potentials have been
found in the context of higher spin gravity \cite{Bunster:2014mua},
\cite{Gary:2014ppa}, \cite{Matulich:2014hea}, hypergravity \cite{Henneaux:2015ywa},
\cite{Henneaux:2015tar} and extended supergravity \cite{Fuentealba:2017fck},
where the coefficients in front of each term in the entropy formula
turn out to be the spin (conformal weight) of the corresponding generator.}
\begin{equation}
S=2N_{\infty}M-2N_{\infty}^{\phi}J-\Phi Q_{e}\,.\label{eq:Smarr1}
\end{equation}
Once the regularity conditions are taking into account it is possible
to identify $N_{\infty}\equiv\beta$, $N_{\infty}^{\phi}\equiv\beta\Omega$
and $\Phi\equiv\beta\Phi_{e}$, where $\beta$ is the inverse of the
Hawking temperature, while $\Omega$ and $\Phi_{e}$ are the chemical
potentials thermodynamically conjugated to the angular momentum $J$
and the electric charge $Q_{e}$, respectively. Thus, the Smarr formula
reads

\begin{equation}
\begin{aligned}M & =\frac{1}{2}TS+\Omega J+\frac{1}{2}\Phi_{e}Q_{e}\end{aligned}
\;,\label{Smarr2}
\end{equation}

It is reassuring to verify that the charged BTZ black hole does satisfy
the Smarr formula \eqref{Smarr2}, as long as the special set of holographic
boundary conditions, determined by \eqref{eq:Euler theorem2}, is
satisfied. In order to carry out this computation one has to consider
the explicit form of the global charges for the black hole, obtained
in \cite{Perez:2015kea}, which are given by
\begin{eqnarray}
M & = & \frac{\pi r_{+}^{2}}{\kappa l^{2}}\left(\frac{1+\omega^{2}}{1-\omega^{2}}\right)-\frac{q_{t}^{2}}{4\pi}\left(\omega^{2}+\left(1+\omega^{2}\right)\log\left(\frac{r_{+}}{l}\right)\right)+\frac{1}{l}\left(\mathcal{V}-q_{\phi}\frac{\delta\mathcal{V}}{\delta q_{\phi}}\right)\ ,\label{rotating masa full}\\
J & = & \frac{2\pi r_{+}^{2}\omega}{\kappa l\left(1-\omega^{2}\right)}-\frac{q_{t}^{2}\omega l}{4\pi}\left(1+\log\left(\frac{r_{+}^{2}}{l^{2}}\right)\right)+q_{t}\frac{\delta\mathcal{V}}{\delta q_{\phi}}\ ,\label{rotating J full}\\
Q_{e} & = & q_{t}\,.
\end{eqnarray}
with $\omega=-q_{\phi}/q_{t}$. The Hawking temperature and the chemical
potentials are
\begin{align}
T & =\frac{\sqrt{1-\omega^{2}}}{2\pi l^{2}r_{+}}\left(r_{+}^{2}-\frac{\kappa l^{2}}{8\pi^{2}}q_{t}^{2}\left(1-\omega^{2}\right)\right)\ ,\\
\Omega & =\frac{\omega}{l}\ ,\\
\Phi_{e} & =-\frac{q_{t}}{2\pi}\left(1-\omega^{2}\right)\log\left(\frac{r_{+}}{l}\right)-\frac{\omega}{l}\frac{\delta\mathcal{V}}{\delta q_{\phi}}+\frac{1}{l}\frac{\delta\mathcal{V}}{\delta q_{t}}\ ,
\end{align}
where the entropy is explicitly given by $S=\frac{4\pi^{2}r_{+}}{\kappa\sqrt{1-\omega^{2}}}$.

\section{Smarr formula from the Euler theorem \label{Euler}}

In this section we will show that by virtue of the special set of
holographic boundary conditions determined by \eqref{eq:Euler theorem2},
it is possible to obtain a homogeneous transformation law for the
extensive quantities, allowing to use the Euler theorem in order to
recover the Smarr formula \eqref{Smarr2} along the lines of its original
derivation given in \cite{Smarr:1972kt} for the Kerr-Newman black
hole. 

Let us consider the scale transformations for the coefficients of
the fall-off \eqref{fall-off} appearing in the energy \eqref{generic M}
and angular momentum \eqref{generic J}, which are given by
\begin{equation}
\begin{aligned}\bar{f}_{\mathcal{F}} & =\lambda^{2}\left[f_{\mathcal{F}}-\left(2h_{R}+\frac{1}{4\pi}\left(q_{t}^{2}+q_{\phi}^{2}\right)\right)\log\left(\lambda\right)\right]\;,\\
\bar{f}_{\mathcal{R}} & =\lambda^{2}\left[f_{\mathcal{R}}+2h_{\mathcal{R}}\log\left(\lambda\right)\right]\;,\\
\bar{j} & =\lambda^{2}\left[j+\frac{l}{2\pi}q_{t}q_{\phi}\log\left(\lambda\right)\right]\;,\\
\bar{h}_{\mathcal{R}} & =\lambda^{2}h_{\mathcal{R}}\;.
\end{aligned}
\label{eq:Scale trafos coeff}
\end{equation}
In the case of a generic $\mathcal{V}$, the mass \eqref{generic M},
the angular momentum \eqref{generic J} and the electric charge \eqref{Electric charge}
transform as
\begin{eqnarray}
\bar{M} & = & \lambda^{2}\left(f_{\mathcal{R}}+f_{\mathcal{F}}+h_{\mathcal{R}}\right)-\frac{1}{4\pi}\left(q_{t}^{2}+q_{\phi}^{2}\right)\lambda^{2}\log\left(\lambda\right)+\frac{1}{l}\left(\bar{\mathcal{V}}-q_{\phi}\frac{\delta\bar{\mathcal{V}}}{\delta q_{\phi}}\right)\ ,\label{eq:barM}\\
\bar{J} & = & \lambda^{2}\left(j+\frac{l}{4\pi}q_{t}q_{\phi}\right)+\frac{l}{2\pi}q_{t}q_{\phi}\lambda^{2}\log\left(\lambda\right)+q_{t}\frac{\delta\bar{\mathcal{V}}}{\delta q_{\phi}}\,,\label{eq:barJ}\\
\bar{Q}_{e} & = & \lambda Q_{e}\,.\label{eq:barQ}
\end{eqnarray}
Note that the electric charge already transforms as homogeneous function
of degree one, while the energy and the angular momentum possess anomalous
scale transformation laws. 

By implementing the holographic boundary conditions, one gets that
the scale transformation of the function $\mathcal{V}$ inherited
from \eqref{eq:Transphi} is given by
\begin{equation}
\bar{\mathcal{V}}\left(\lambda q_{\mu}\right)=\lambda^{2}\left[\mathcal{V}+\frac{l}{4\pi}\left(q_{t}^{2}-q_{\phi}^{2}\right)\log\left(\lambda\right)\right]\;.\label{eq:Trafo Nu-1}
\end{equation}
Remarkably, by replacing \eqref{eq:Trafo Nu-1} into \eqref{eq:barM}
and \eqref{eq:barJ}, the additional anomalous logarithmic terms precisely
cancel out, such that the energy and the angular momentum now transform
as homogeneous functions of degree two, i.e.,
\begin{equation}
\bar{M}=\lambda^{2}M\quad,\quad\bar{J}=\lambda^{2}J\,.
\end{equation}
 where $M$ and $J$ are given in \eqref{generic M} and \eqref{generic J},
respectively.

At this point, we focus in the charged BTZ black hole, whose entropy,
given by $S=\frac{4\pi^{2}}{\kappa}\mathcal{R}(r_{+})$, transforms
as a homogeneous function of degree one, i.e., $\bar{S}=\lambda S$.
In consequence, the entropy is a homogeneous function of degree $\frac{1}{2}$
in $\left(M,J,Q_{e}^{2}\right)$, i.e.,
\begin{equation}
\begin{aligned}S\left(\sigma M,\sigma J,\sigma Q_{e}^{2}\right) & =\sigma^{1/2}S\left(M,J,Q_{e}^{2}\right)\;.\end{aligned}
\label{Homogenous Mass}
\end{equation}
Hence, by direct application of the Euler theorem for homogeneous
functions, the above equation yields
\begin{equation}
\frac{1}{2}S=M\left.\frac{\partial S}{\partial M}\right|_{J,Q_{e}}+J\left.\frac{\partial S}{\partial J}\right|_{M,Q_{e}}+\frac{1}{2}Q_{e}\left.\frac{\partial S}{\partial Q_{e}}\right|_{M,J}\,,\label{eq:Smarr2.5}
\end{equation}
and by virtue of the first law $\delta S=N_{\infty}\delta M-N_{\infty}^{\phi}\delta J-\Phi\delta Q_{e}$;
\begin{equation}
N_{\infty}=\left.\frac{\partial S}{\partial M}\right|_{J,Q_{e}}\quad,\quad N_{\infty}^{\phi}=-\left.\frac{\partial S}{\partial J}\right|_{M,Q_{e}}\quad,\quad\Phi=-\left.\frac{\partial S}{\partial Q_{e}}\right|_{M,J}\,,
\end{equation}
the equation \eqref{eq:Smarr2.5} can be cast directly to the expected
Smarr formula 
\begin{equation}
S=2N_{\infty}M-2N_{\infty}^{\phi}J-\Phi Q_{e}\,,\label{eq:Smarr3}
\end{equation}
provided $N_{\infty}=\beta$, $N_{\infty}^{\phi}=\beta\Omega$ and
$\Phi=\beta\Phi_{e}$.

Note that for the simplest choice of boundary conditions, $\mathcal{V}=0$,
the global charges given by \eqref{generic M}, \eqref{generic J}
correspond to the ones found in \cite{Martinez:1999qi}. In this case,
the logarithmic terms in \eqref{eq:Scale trafos coeff} lead to non-homogeneous
contributions in the scale transformations of the global charges \eqref{generic M},
\eqref{generic J}, as it can be seen explicitly
\begin{eqnarray}
\bar{M} & = & \lambda^{2}M-\frac{1}{4\pi}\left(q_{t}^{2}+q_{\phi}^{2}\right)\lambda^{2}\log\left(\lambda\right)\ ,\label{eq:barM-1}\\
\bar{J} & = & \lambda^{2}J+\frac{l}{2\pi}q_{t}q_{\phi}\lambda^{2}\log\left(\lambda\right)\,,\label{eq:barJ-1}
\end{eqnarray}
yielding, by means of the Euler theorem, the relation \eqref{eq:SmarrSpoiled}
\begin{equation}
M=\frac{1}{2}TS+\Omega J+\frac{1}{2}\Phi_{e}Q_{e}+\frac{1}{8\pi}\left(1-l^{2}\Omega^{2}\right)Q_{e}^{2}\,.\label{eq:SmarrSpoiled-1}
\end{equation}

Therefore, the logarithmic terms in \eqref{eq:barM-1} and \eqref{eq:barJ-1}
preclude the possibility to obtain the Smarr formula \eqref{Smarr2},
since the assumption of the homogeneity scaling property of the global
charges are no longer satisfied.

\section{Ending remarks \label{Concludings}}

In this work it has been shown that the Smarr fomula for the charged
BTZ black hole emerges from two different approaches. Both of them
are based on the preservation of the fall-off of the fields under
scale transformations which leave the reduced action principle invariant.
In the first approach, we have proved that the scale invariance of
the theory for stationary and circularly symmetric configurations
is associated to a radially conserved charge. This conservation law
leads to the Smarr formula as long as a special set of holographic
boundary conditions is fulfilled. In the second approach, it was found
that the same set of holographic boundary conditions confers the homogeneity
scaling property to the global charges, allowing to derive the Smarr
formula of the black hole through the Euler theorem.

Throughout this work, we have considered that the cosmological constant
is a coupling constant fixed without variation. Nonetheless, the problem
related to the spoil of the homogeneity property in the extensive
variables of the electrically charged BTZ black hole has been addressed
at some extent under the rescaling of the cosmological constant, which
leads to the introduction of additional thermodynamical terms both
in the first law and in the energy formula \eqref{eq:SmarrSpoiled}
(see e.g. \cite{LarranagaRubio:2007fly}, \cite{Dehghani:2016agl}).
This treatment might be consistently carried out promoting the cosmological
constant to a canonical variable through the mechanism described in
\cite{Henneaux:1984ji}. However, recent results has shown the existence
of a superselection rule that forbids a superposition of quantum states
with different values of the cosmological constant in three dimensions
\cite{Bunster:2014cna}, so that its value would be definite, and
in consequence, it cannot be rescaled.

Finally, it is noteworthy that apart of obtaining the Smarr formula
for the charged BTZ black hole, the set of holographic boundary conditions
endowed with the additional requirement of Lorentz symmetry makes
the energy spectrum of the black hole nonnegative, and the electric
charge bounded from above, for a fixed value of the energy \cite{Perez:2015kea}.
This strongly suggests that the solution might be stable, so it would
be interesting to carry out a thermodynamical analysis of the stability
of the charged BTZ black hole by considering a generic set of boundary
conditions.

\acknowledgments

We thank Marcela Cárdenas, Ernesto Frodden, Alfredo Pérez and David
Tempo for helpful comments and discussions. Special thanks to Ricardo
Troncoso for his careful reading of the manuscript and encouragement.
C.E. thanks Conicyt through Becas Chile programme for financial support.
This research has been partially supported by Fondecyt grants Nº 3170707,
3170772. Centro de Estudios Científicos (CECs) is funded by the Chilean
Government through the Centers of Excellence Base Financing Program
of Conicyt.


\begin{thebibliography}{10}
\bibitem{Smarr:1972kt}L.~Smarr,   ``Mass formula for Kerr black holes,''   Phys.\ Rev.\ Lett.\  {\bf 30}, 71 (1973)   Erratum: [Phys.\ Rev.\ Lett.\  {\bf 30}, 521 (1973)].     

\bibitem{Banados:1992wn}M.~Banados, C.~Teitelboim and J.~Zanelli,   ``The Black hole in three-dimensional space-time,''   Phys.\ Rev.\ Lett.\  {\bf 69}, 1849 (1992)      [hep-th/9204099].   

\bibitem{Banados:1992gq}M.~Banados, M.~Henneaux, C.~Teitelboim and J.~Zanelli,   ``Geometry of the (2+1) black hole,''   Phys.\ Rev.\ D {\bf 48}, 1506 (1993)   Erratum: [Phys.\ Rev.\ D {\bf 88}, 069902 (2013)]     [gr-qc/9302012].   

\bibitem{Strominger:1997eq}A.~Strominger,   ``Black hole entropy from near horizon microstates,''   JHEP {\bf 9802}, 009 (1998)     [hep-th/9712251].   

\bibitem{Brown:1986nw}J.~D.~Brown and M.~Henneaux,   ``Central Charges in the Canonical Realization of Asymptotic Symmetries: An Example from Three-Dimensional Gravity,''   Commun.\ Math.\ Phys.\  {\bf 104}, 207 (1986).     

\bibitem{Cai:1996df}R.~G.~Cai, Z.~J.~Lu and Y.~Z.~Zhang,   ``Critical behavior in (2+1)-dimensional black holes,''   Phys.\ Rev.\ D {\bf 55}, 853 (1997)     [gr-qc/9702032].   

\bibitem{Clement:2003sr}G.~Clement,   ``Black hole mass and angular momentum in 2+1 gravity,''   Phys.\ Rev.\ D {\bf 68}, 024032 (2003)    [gr-qc/0301129].   

\bibitem{Perez:2015kea}A.~Pérez, M.~Riquelme, D.~Tempo and R.~Troncoso,   ``Conserved charges and black holes in the Einstein-Maxwell theory on AdS$_{3}$ reconsidered,''   JHEP {\bf 1510}, 161 (2015)     [arXiv:1509.01750 [hep-th]].   

\bibitem{Perez:2015jxn}A.~Pérez, M.~Riquelme, D.~Tempo and R.~Troncoso,   ``Asymptotic structure of the Einstein-Maxwell theory on AdS$_{3}$,''   JHEP {\bf 1602}, 015 (2016)     [arXiv:1512.01576 [hep-th]].   

\bibitem{Clement:1993kc}G.~Clement,   ``Classical solutions in three-dimensional Einstein-Maxwell cosmological gravity,''   Class.\ Quant.\ Grav.\  {\bf 10}, L49 (1993).   

\bibitem{Martinez:1999qi}C.~Martinez, C.~Teitelboim and J.~Zanelli,   ``Charged rotating black hole in three space-time dimensions,''   Phys.\ Rev.\ D {\bf 61}, 104013 (2000)     [hep-th/9912259].   

\bibitem{Bravo-Gaete:2015iwa}M.~Bravo-Gaete, S.~Gomez and M.~Hassaine,   ``Cardy formula for charged black holes with anisotropic scaling,''   Phys.\ Rev.\ D {\bf 92}, no. 12, 124002 (2015)     [arXiv:1510.04084 [hep-th]].   

\bibitem{Banados:2005hm}M.~Banados and S.~Theisen,   ``Scale invariant hairy black holes,''   Phys.\ Rev.\ D {\bf 72}, 064019 (2005)    [hep-th/0506025].   

\bibitem{Deser:1985pk}S.~Deser and P.~O.~Mazur,   ``Static Solutions in $D=3$ Einstein-maxwell Theory,''   Class.\ Quant.\ Grav.\  {\bf 2}, L51 (1985).      

\bibitem{Gott:1986bp}J.~R.~Gott, J.~Z.~Simon and M.~Alpert,   ``General Relativity in a (2+1)-dimensional Space-time: An Electrically Charged Solution,''   Gen.\ Rel.\ Grav.\  {\bf 18}, 1019 (1986).   

\bibitem{Peldan:1992mp}P.~Peldan,   ``Unification of gravity and Yang-Mills theory in (2+1)-dimensions,''   Nucl.\ Phys.\ B {\bf 395}, 239 (1993)      [gr-qc/9211014].   

\bibitem{Kamata:1995zu}M.~Kamata and T.~Koikawa,   ``The Electrically charged BTZ black hole with self (antiself) dual Maxwell field,''   Phys.\ Lett.\ B {\bf 353}, 196 (1995)    [hep-th/9505037].   

\bibitem{Chan:1995uh}K.~C.~K.~Chan,   ``Comment on the calculation of the angular momentum for the (anti)selfdual charged spinning BTZ black hole,''   Phys.\ Lett.\ B {\bf 373}, 296 (1996)     [gr-qc/9509032].   

\bibitem{Clement:1995zt}G.~Clement,   ``Spinning charged BTZ black holes and selfdual particle - like solutions,''   Phys.\ Lett.\ B {\bf 367}, 70 (1996)     [gr-qc/9510025].   

\bibitem{Hirschmann:1995he}E.~W.~Hirschmann and D.~L.~Welch,   ``Magnetic solutions to (2+1) gravity,''   Phys.\ Rev.\ D {\bf 53}, 5579 (1996)     [hep-th/9510181].   

\bibitem{Cataldo:1996ue}M.~Cataldo and P.~Salgado,   ``Static Einstein-Maxwell solutions in (2+1)-dimensions,''   Phys.\ Rev.\ D {\bf 54}, 2971 (1996).    

\bibitem{Kamata:1996vg}M.~Kamata and T.~Koikawa,   ``(2+1)-dimensional charged black hole with (anti-)selfdual Maxwell fields,''   Phys.\ Lett.\ B {\bf 391}, 87 (1997)     [hep-th/9605114].   

\bibitem{Cataldo:1999fh}M.~Cataldo and P.~Salgado,   ``Three dimensional extreme black hole with self (anti-self) dual Maxwell field,''   Phys.\ Lett.\ B {\bf 448}, 20 (1999).      

\bibitem{Cataldo:2002fh}M.~Cataldo,   ``Azimuthal electric field in a static rotationally symmetric (2+1)-dimensional space-time,''   Phys.\ Lett.\ B {\bf 529}, 143 (2002)     [gr-qc/0201047].   

\bibitem{Dias:2002ps}O.~J.~C.~Dias and J.~P.~S.~Lemos,   ``Rotating magnetic solution in three-dimensional Einstein gravity,''   JHEP {\bf 0201}, 006 (2002)      [hep-th/0201058].   

\bibitem{Cataldo:2004uw}M.~Cataldo, J.~Crisostomo, S.~del Campo and P.~Salgado,   ``On magnetic solution to (2+1) Einstein-Maxwell gravity,''   Phys.\ Lett.\ B {\bf 584}, 123 (2004)    [hep-th/0401189].   

\bibitem{Matyjasek:2004pg}J.~Matyjasek and O.~B.~Zaslavskii,   ``Extremal limit for charged and rotating (2+1)-dimensional black holes and Bertotti-Robinson geometry,''   Class.\ Quant.\ Grav.\  {\bf 21}, 4283 (2004)      [gr-qc/0404090].   

\bibitem{Garcia-Diaz:2013baa}A.~A.~Garcia-Diaz,   ``Three dimensional stationary cyclic symmetric Einstein-Maxwell solutions; black holes,''   Annals of Physics {\bf {\bf 324}}, 2004 (2009)    [arXiv:1307.6655 [gr-qc]].   

\bibitem{AyonBeato:2004qr}E.~Ayon-Beato, M.~Cataldo and A.~A.~Garcia,   ``Electromagnetic fields in stationary cyclic symmetric 2+1 gravity,''   

\bibitem{Benguria:1976in}R.~Benguria, P.~Cordero and C.~Teitelboim,   ``Aspects of the Hamiltonian Dynamics of Interacting Gravitational Gauge and Higgs Fields with Applications to Spherical Symmetry,''   Nucl.\ Phys.\ B {\bf 122}, 61 (1977).    

\bibitem{Regge:1974zd}T.~Regge and C.~Teitelboim,   ``Role of Surface Integrals in the Hamiltonian Formulation of General Relativity,''   Annals Phys.\  {\bf 88}, 286 (1974).   

\bibitem{Horowitz:1999jd}G.~T.~Horowitz and V.~E.~Hubeny,   ``Quasinormal modes of AdS black holes and the approach to thermal equilibrium,''   Phys.\ Rev.\ D {\bf 62}, 024027 (2000)    [hep-th/9909056].   

\bibitem{Bunster:2014mua}C.~Bunster, M.~Henneaux, A.~Pérez, D.~Tempo and R.~Troncoso,   ``Generalized Black Holes in Three-dimensional Spacetime,''   JHEP {\bf 1405}, 031 (2014)     [arXiv:1404.3305 [hep-th]].   

\bibitem{Gary:2014ppa}M.~Gary, D.~Grumiller, M.~Riegler and J.~Rosseel,   ``Flat space (higher spin) gravity with chemical potentials,''   JHEP {\bf 1501}, 152 (2015)      [arXiv:1411.3728 [hep-th]].   

\bibitem{Matulich:2014hea}J.~Matulich, A.~Perez, D.~Tempo and R.~Troncoso,   ``Higher spin extension of cosmological spacetimes in 3D: asymptotically flat behaviour with chemical potentials and thermodynamics,''   JHEP {\bf 1505}, 025 (2015)      [arXiv:1412.1464 [hep-th]].   

\bibitem{Henneaux:2015ywa}M.~Henneaux, A.~Perez, D.~Tempo and R.~Troncoso,   ``Hypersymmetry bounds and three-dimensional higher-spin black holes,''   JHEP {\bf 1508}, 021 (2015)     [arXiv:1506.01847 [hep-th]].   

\bibitem{Henneaux:2015tar}M.~Henneaux, A.~Pérez, D.~Tempo and R.~Troncoso,   ``Extended anti-de Sitter Hypergravity in $2+1$ Dimensions and Hypersymmetry Bounds,''    arXiv:1512.08603 [hep-th].   

\bibitem{Fuentealba:2017fck}O.~Fuentealba, J.~Matulich and R.~Troncoso,   ``Asymptotic structure of $\mathcal{N}=2$ supergravity in 3D: extended super-BMS$_3$ and nonlinear energy bounds,''   JHEP {\bf 1709}, 030 (2017)    [arXiv:1706.07542 [hep-th]].   

\bibitem{LarranagaRubio:2007fly}E.~A.~Larranaga Rubio,   ``On the first law of thermodynamics for (2+1) dimensional charged BTZ black hole and charged de Sitter space,''   arXiv:0707.2256 [gr-qc].   

\bibitem{Dehghani:2016agl}M.~Dehghani,   ``Thermodynamics of $(2+1)$-dimensional charged black holes with power-law Maxwell field,''   Phys.\ Rev.\ D {\bf 94}, no. 10, 104071 (2016).    

\bibitem{Henneaux:1984ji}M.~Henneaux and C.~Teitelboim,   ``The Cosmological Constant As A Canonical Variable,''   Phys.\ Lett.\  {\bf 143B}, 415 (1984).   

\bibitem{Bunster:2014cna}C.~Bunster and A.~Pérez,   ``Superselection rule for the cosmological constant in three-dimensional spacetime,''   Phys.\ Rev.\ D {\bf 91}, no. 2, 024029 (2015)    [arXiv:1412.1492 [hep-th]].   
\end{thebibliography}
\end{document}